\documentclass[doublecol]{epl2}
\usepackage{graphicx}
\usepackage{amssymb,amsmath}

\newcommand{\n}{\overline{n}_L}
\newcommand{\vn}{\sigma^2_{n_L}}
\newcommand{\Tr}{\mathrm{Tr}}
\newcommand{\bino}[2]{\frac{#1!}{#2!(#1-#2)!}}
\newcommand{\Rb}{$^{87}$Rb }
\newcommand{\nK}{\;\mathrm{nK}\cdot k_B}

\title{Dynamical Realization of Macroscopic Superposition States of Cold Bosons in a Tilted Double Well}

\author{L.~D. Carr, D.~R. Dounas-Frazer, and M.~A. Garcia-March}
\institute{Department of Physics, Colorado School of Mines, Golden,
  CO, 80401}
\date{\today}

\abstract{
We present exact expressions for the quantum sloshing of Bose-Einstein condensates in a tilted two-well potential.  Tunneling is suppressed by a
small potential difference between wells, or tilt. However,
tunneling resonances occur for critical values of the tilt when the
barrier is high.  At resonance, tunneling times on the order of
10-100 ms are possible. Furthermore, such tilted resonances lead to
a dynamical scheme for creating few-body NOON-like macroscopic superposition states which are protected
by the many body wavefunction against potential fluctuations.
}

\pacs{03.75.Lm}{Tunneling, Bose-Einstein condensates in periodic potentials}
\begin{document}

\maketitle

Bose-Einstein condensates (BECs) in optical lattices are an ideal
medium for studying a vast range of quantum many-body phenomena~\cite{lewensteinM2007,bloch2008},
including macroscopic quantum tunneling.  A two-well potential is a
simple limiting case which nevertheless exhibits rich quantum
behavior.  Spatially separate BECs in a two-well potential have
been created in experiments~\cite{Shin:2005} and tunneling
times on the order of 50 ms have been observed~\cite{Albiez:2005}.
The lifetime of a typical experiment is 1-100~s~\cite{Tolra:2004}.  This system, in certain limits (see Eq.~(\ref{eqn:condition}) below), maps onto the Lipkin-Meshkov-Glick (LMG) model~\cite{lipkinHJ1965,vidalJ2004},\footnote{Sometimes called ``two-mode Bose Hubbard'' or ``two-site Bose Hubbard'' in cold
  quantum gases literature, although the LMG model precedes the Bose-Hubbard
  model by 24 years.} the $N$-body generalization of the two-state problem, in which $N$ particles can occupy two single-particle modes.  Dynamical instability at the level of mean field theory (MFT) is associated with generation of strongly correlated states on a microscopic exact diagonalization (ED) level~\cite{Dagnino:2009}; the LMG model can be solved by both MFT and ED.  BECs offer the exciting possibility of realizing macroscopic superposition (MS) states (NOON states) with tens to thousands of particles, and thereby pushing the limits of quantum mechanics~\cite{leggett2001}.  However, the physical context of a BEC in a double well has quite different regimes.  Thus different approaches are required, including MFT~\cite{Milburn:1997,Smerzi:1997,Zapata:1998}, the multiconfigurational time-dependent Hartree (MCTDH) theory~\cite{masiello2006,Alon:2008}, and our own~\cite{Dounas-Frazer:2007} and others'~\cite{watanabeG2010,weissC2008} work on ED of the LMG model with additional terms, including \emph{tilt} or bias.  Without the active use of tilt, MS states are destroyed by potential fluctuations~\cite{Dounas-Frazer:2007,Milburn:1997}.

In this Letter, we use a biased LMG model to investigate the
\emph{quantum sloshing} of many bosons in a tilted double-well, with the goal of guiding dynamical creation of MS states in BEC double-well experiments.  Quantum sloshing is the tunneling dynamics of a system in which all atoms are initially localized in one well.  At the quarter and three-quarter periods of the ensuing cyclical dynamics, one finds an MS state. The basic concept of quantum sloshing is similar to what occurs in an rf SQUID, in which the many body wavefunction oscillates between two macroscopically distinct states~\cite{friedman2000}.

The LMG model is applicable for
\begin{equation}
\chi \equiv [(N^2-1) U]/ (2\hbar\omega) \lesssim 1 \,,
\label{eqn:condition}
\end{equation}
where $N$ is the number of atoms, $U$ is the interaction energy, and $\omega$ is the local trap frequency in each of the two wells~\cite{Dounas-Frazer:2007,Dounas-Frazerthesis:2006}.  Criterion~(\ref{eqn:condition}) means that only the lowest single-particle state in either well is occupied.  Additionally, it is required that all dynamical perturbation have an energy much less than $\hbar\omega$.  Under these restrictions, there are two regimes: $\zeta/N\equiv J/N|U| \gg 1$, where $J$ is the tunneling energy, we denote the \emph{Josephson regime}; $\zeta \ll 1$ we denote the \emph{Fock regime}~\cite{leggett2001}.  It is in the latter that MS states occur.
The experimentally observed phenomenon of self-trapping, in which the system becomes stuck in one well, can be attributed to long tunneling times~\cite{Salgueiro:2006}.  These oscillations with long tunneling times are associated with the presence of Schr\"odinger-cat-like, or NOON-like, MS states.  Such states collapse to a narrow distribution of Fock states in the presence of a very small tilt~\cite{Dounas-Frazer:2007}, and tunneling is exponentially suppressed.
Past studies on the suppression of tunneling have focused on environmental effects such as finite temperature or coupling to a reservoir~\cite{Paz:1993,pitaevskii2001,Antunes:2006}. Therefore, in the presence of a small tilt, finite temperature, or coupling to a reservoir, the MS states are dynamically unaccessible.
Tilt displays radically different behavior than these other forms of suppression of the tunneling. Namely, \emph{tunneling resonances} occur for critical values of the tilt when the barrier is high.  At resonance, MS states states reappear~\cite{Dounas-Frazer:2007},  and therefore tunneling is again observed.  We show that at tunneling resonances the oscillation time between wells is hundreds of orders of magnitude faster and less sensitive
to deviations in the tilt than in the symmetric case.

This speed-up permits us to propose a simple scheme for the creation of
MS states both for
few-body and many-body
systems.  Whereas past proposals involved ramping
the barrier height~\cite{Mahmud:2005} or continuous variation of
atom-atom interactions via Feshbach resonance~\cite{Huang:2006},
MS states are realized periodically in our scheme when all
parameters are fixed.  These MS states take the form of \textit{protected} NOON states.  Using the tunneling times of quantum sloshing to positively identify NOON or NOON-like states has been a significant tool in the study of SQUID-based NOON states, where the double well consists of right- and left-circulating states on a ring~\cite{friedman2000}. In cold atom experiments, quantum tunneling effects have so far been restricted to one or two atoms~\cite{anderlini2007,foelling2007}, and MS states of NOON-like form have yet to be clearly identified.

We briefly mention methods for regimes other than that of the biased LMG.  For $\chi \gg 1$, MFT is applicable for $\zeta/N \gg 1$; in this regime, called the ``linear'' or ``Josephson'' regime~\cite{leggett2001},\footnote{Sols and Leggett subdivide this regime into ``Rabi'' and ``Josephson''; we use
  only the term ``Josephson'' for simplicity.} the double-well system can be described by a pendulum in a 2D phase-number phase space~\cite{Javanainen:1987,Milburn:1997, Smerzi:1997, Zapata:1998, Mahmud:2005,  Ananikian:2006}.  In contrast, for $\zeta/N \gtrsim 1$, called the ``nonlinear'' or ``self-trapping'' regime, MFT methods find macroscopic self trapping in one well~\cite{Milburn:1997,Smerzi:1997, Zapata:1998, Ananikian:2006}.  The concept of phase is well-defined in the Josephson regime, together with a clear semiclassical limit; it is not well-defined in the ``nonlinear'' regime~\cite{Zapata:1998, Kohler:2001}, and then requires a more strongly quantum approach~\cite{Paraoanu:2001,Leggett:1991}.

MCTDH theory is such an approach. It does indeed find different dynamics from mean-field predictions, particularly in the Fock regime~\cite{Sakmann:2009}.  This purely numerical method is more exact~\cite{Alon:2008} than MFT or LMG approaches~\cite{Gati:2006, Huang:2006, Mahmud:2005, Jaaskelainen:2005, Salgueiro:2006,Dounas-Frazer:2007} and in principle superior, as it can span all regimes.  However, our modified LMG approach has the advantage that it is simpler and leads to exact and perturbative analytical expressions together with straightforward simulations; a mix of analytical and numerical methods at different levels of approximation is useful.  Moreover, because our methods are not computationally intensive they can be easily extended to two and three dimensions~\cite{Dounas-Frazerthesis:2006}, unlike the more exact but computationally demanding MCTDH theory.  Under condition~(\ref{eqn:condition}), MCTDH and LMG methods should converge.

Thus we work with the biased LMG model.  Although mean-field theory for the asymmetric trap has been considered before~\cite{Smerzi:1997}, the MFT approximation is, as we have said, a poor one in the Fock regime, $\zeta \lesssim 1$.  Experimentally, tilt appears both as a
systematic error and deliberately in device
applications~\cite{Ferrari:2006,
  Hall:2006, anderlini2007,foelling2007,Seaman:2006}. Tilted optical
lattices are especially relevant to applications in
gravitometry~\cite{Ferrari:2006}, quantum
computing~\cite{Brennen:1999,anderlini2007}, and atomtronics~\cite{Seaman:2006}.

The two-mode Hamiltonian for $N$ weakly interacting bosons in a
tilted two-well potential, or biased LMG model, is
\begin{equation}\label{eq:1}
  \hat{H} = -J\sum_{j\neq j'}\hat{b}^{\dagger}_j\hat{b}_{j'}
  + U\sum_j\hat{n}_j\left(\hat{n}_j-1\right)+\Delta V\hat{n}_L,
\end{equation}
where the subscript $j\in\{L,R\}$ is the well or site index, $J$ is
the hopping strength, $U$ is the interaction potential, and $\Delta
V$ is the tilt. Here $\hat{b}_j$ and $\hat{b}_j^{\dagger}$ satisfy
the usual bosonic annihilation and creation commutation relations
and $\hat{n}_j\equiv\hat{b}_j^{\dagger}\hat{b}_j$. Eq. (\ref{eq:1})
can be derived from first-principles quantum field theory for weakly
interacting bosons at zero temperature.   An arbitrary state vector in Fock
space is given by
\begin{equation}
|\psi\rangle =
\textstyle\sum_{n_L=0}^Nc_{n_L}|n_L,\;N-n_L\rangle\,,
\end{equation} where $n_L$ is the number
of particles in the left well and $c_{n_L}\in\mathbb{C}$.  We require the total number of
particles $N$ to be constant.  Under this restriction, the
Hamiltonian reduces to an $(N+1)\times(N+1)$ tridiagonal
matrix~\cite{Mahmud:2005}.  We consider the
dynamics of a system in which all particles initially occupy the
right well, i.e., $|\psi(t=0)\rangle = |0,\,N\rangle$. In the
Schr\"odinger picture, the time evolved ket is
$|\psi(t)\rangle\equiv\exp(-i\hat{H}t/ \hbar)|\psi\rangle$.  The
probability of finding $n_L$ particles in the left well at some time
$t>0$ is $P_{n_L}(t) \equiv |\langle n_L,\,N-n_L|\psi(t)\rangle|^2$,
the average occupation of the left well is $\n(t) \equiv
\langle\psi(t)|\hat{n}_L|\psi(t)\rangle$, and the average variance is $\vn(t) \equiv
\langle\psi(t)|\hat{n}_L^2|\psi(t)\rangle-\n^2$.

\begin{figure}\center
\includegraphics[width=3.25in]{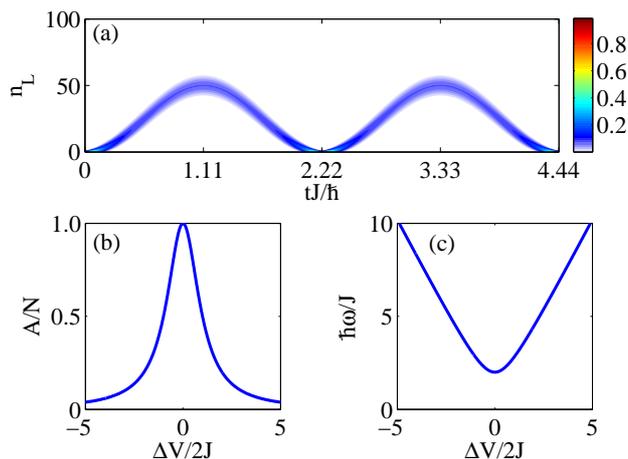}
\caption{(Color online) {\it Suppression of tunneling for noninteracting atoms.} (a)
  Shown are the probability densities $P_{n_L}(t)$ (colorbar) for all number states when
  $N=100$, $U=0$, and $\Delta V=2J$. Only
  $N/2=50$ particles tunnel between wells.  (b) The tunneling amplitude and
  (c) the frequency of oscillation as a
  function of tilt.  When $\Delta V > 2J\sqrt{N-1}$, tunneling is completely
  suppressed.
  Particles tunnel between wells faster in a tilted potential than in
  a symmetric potential.
\label{fig:1}}
\end{figure}

We first consider the simple case of noninteracting particles, $U=0$, in
a symmetric potential, $\Delta V=0$, to illustrate the problem. This case is exactly solvable.  The
probability of finding all particles in the right well, i.e., $n_L=0$,
is
\begin{equation}
  P_0(t)=\cos^{2N}(Jt/\hbar).
\end{equation}
The tunneling period is $T \equiv \pi\hbar/J$, which is independent
of $N$.  When $t = T/2$, the system is in state $|N,\,0\rangle$ and
all particles have tunneled into the left well.  The average
occupation and variance of the left well are
\begin{eqnarray}
  \n(t) &=& N\sin^2(Jt/\hbar),\\
  \vn(t) &=& (N/4)\sin^2(2Jt/\hbar).
\end{eqnarray}
The particles therefore tunnel sinusoidally between wells with a
frequency $2J/\hbar$.  The variance is greatest when $t=T/4$.  At
this time, the probability of finding $n_L$ particles in the left well
is
\begin{equation}\label{eq:coherent}
  P_{n_L}(T/4) = 2^{-N}N!/[n_L!(N-n_L)!].
\end{equation}
The system is in a {\it truncated coherent state}, i.e., a binomial
superposition of all number-states.

The tilted case is also straightforward~\cite{Dounas-Frazerthesis:2006}, although to our knowledge the following expressions have not yet appeared in the literature.  The dynamics are slightly different when $\Delta V\neq0$, since the occupation of the left well now is
\begin{equation}
  \n(t) = A\sin^2(\omega t/2),
\end{equation}
where the amplitude and frequency of oscillation are
\begin{align}
  A &\equiv N/[1+(\Delta V/2J)^2],\label{eq:A}\\
  \omega &\equiv (2J/\hbar)\sqrt{1+(\Delta V/2J)^2}.\label{eq:w}
\end{align}
When $\Delta V=2J$, only $N/2$ particles tunnel between wells.  Fig.
\ref{fig:1}(a) shows the probability densities $P_{n_L}(t)$ in this
case.  In Figs. \ref{fig:1}(b) and \ref{fig:1}(c), Eqs. (\ref{eq:A})
and  (\ref{eq:w}) are plotted as a function of $\Delta V$. From the expression for $A$ we observe that tunneling
between wells is completely suppressed when $|\Delta V| >
2J\sqrt{N-1}$. Because the hopping strength $J$ is much smaller
than the barrier height, tunneling is
highly sensitive to small tilt.

\begin{figure}[t]\center
\includegraphics[width=3.25in]{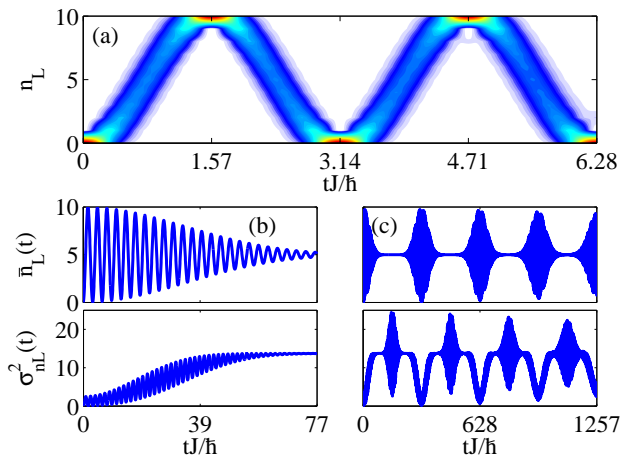}
\caption{(Color online) {\it Damped tunneling in the Josephson regime.} (a) Probability densities $P_{n_L}(t)$ for all number states when
  $N=10$ and $\zeta/N\equiv J/NU =10$ for $t\ll T_{1/2}$.  All particles tunnel between
  wells with period $T=\pi\hbar/J$. (b),(c) Average
  occupation  (top panel) and
  number variance (bottom panel) of the left well for longer
  times.  (b) Oscillations
  between wells are damped by atom-atom interactions.  (c) The first
  tunneling revival occurs when $t=T_r\equiv\pi\hbar/U$.  The colorbar is the
  same as in Fig.~\ref{fig:1}. \label{fig:2}}
\end{figure}

We proceed to consider how a small interaction term changes this scenario in the Josephson regime, $\zeta \gg 1$, in a symmetric potential, $\Delta V=0$.
For the non-interacting system a single frequency $2J/\hbar$ characterizes $\n(t)$; in contrast, $N$ dominant frequencies emerge in the interacting system in the Josephson regime.  The average
occupation of the left well is given by the modulated signal
\begin{equation}
   \n(t) =
   (N/2)\left[1-\cos(2Jt/\hbar)\cos^{N-1}(Ut/\hbar)\right],
   \label{eqn:Nfreq}
\end{equation}
to lowest order in perturbation theory~\cite{Dounas-Frazerthesis:2006} in $N/\zeta$.
We have also verified this result through simulations. In Eq.~(\ref{eqn:Nfreq}) the high frequency carrier depends only on the hopping
strength $J$ while the low frequency envelope depends on both the
interaction  potential $U$ and the total number of particles $N$.  The
envelope reaches half its maximum value when
\begin{equation}\label{eq:T1/2}
 t = T_{1/2} \equiv (\hbar/U)\cos^{-1}[2^{-1/(N-1)}].
\end{equation}
At times $t\ll T_{1/2}$, all particles tunnel between wells with period
$T$, as in Fig. \ref{fig:2}(a).  At times near $T_{1/2}$, on the other
hand, only half the particles tunnel between wells with period $T$.
When $t\simeq2T_{1/2}$, there is essentially no tunneling (see
Fig. \ref{fig:2}(b)).  Small interactions thus damp the oscillations
between wells~\cite{Zapata:2003, Salgueiro:2006}.
However, tunneling revivals occur periodically with
period $T_r \equiv \pi\hbar/U$.  The first tunneling revival occurs when
$|t-T_r|<T_{1/2}$, as shown in Fig. \ref{fig:2}(c).  The separation of
time scales, $T_{1/2}\ll T_r$, occurs only for $N\gg1$, as evident in
Eq. (\ref{eq:T1/2}).

For the remainder of our discussion, we turn to the high barrier
limit, $\zeta \ll 1$, as it is key to the dynamic production of
MS states.
We assume $U>0$ without loss of generality with
respect to the dynamics.  Using perturbation theory, it can be shown~\cite{Dounas-Frazer:2007,Dounas-Frazerthesis:2006} that the eigenstates are MS states of the form $|\phi_{\pm}\rangle \equiv\left(|N-n_L,\,n_L\rangle \pm |n_L,\,N-n_L\rangle\right)/ \sqrt{2}$ to lowest order in $\zeta$. The degenerate number states in the $\zeta=0$ limit split into symmetric and antisymmetric MS states for small $\zeta$, with an energy difference of $\Delta E_{N-n_L}$. The two eigenstates with the highest eigenvalue are nearly-degenerate MS states of the form $|\phi_{\pm}\rangle \equiv\left(|N,\,0\rangle \pm |0,\,N\rangle\right)/ \sqrt{2}$. The energy difference between $|\phi_{\pm}\rangle$ is
\begin{equation}
  \Delta E_N = 4U(\zeta/2)^NN/[(N-1)!].
  \label{eqn:splittingNoTilt}
\end{equation}
The characteristic frequency is $\omega_N = \Delta E_N/\hbar$. Notice that since $\Delta E_N$ is a very small number ($\zeta \ll 1$),  $\omega_N$ is also very small, and decreases rapidly with increasing $N$.

\begin{figure}\center
\includegraphics[width=3.25in]{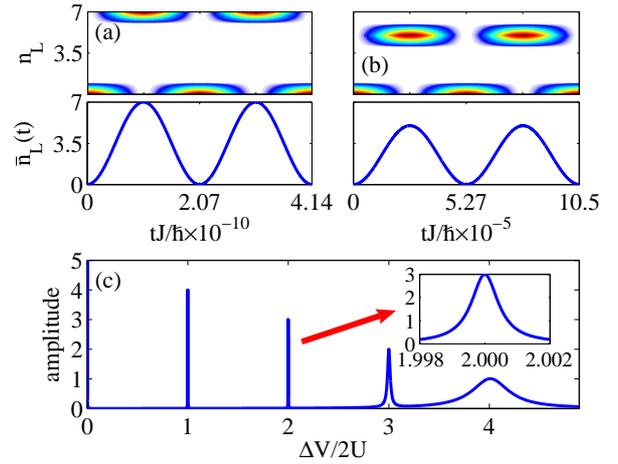}
\caption{(Color online) {\it Tunneling resonances in a few-atom system.} Shown are
  the probability densities $P_{n_L}(t)$ when $N=7$ and $\zeta =
  0.1$, for (a) $\Delta V=0$ and (b) $\Delta V=4U$.  (a) All particles
  tunnel between wells with period $T_N$.  At time $t=T_N/4$, the
  system is described by an MS state. (b) Only $N-2=5$
  particles tunnel between wells.  The oscillation frequency
  is 5 orders of magnitude faster than the symmetric case. (c)
  Tunneling amplitude as a function of tilt $\Delta V$ for
  $N=5$ and $\zeta = 0.1$.  Tunneling resonances occur when
  $\Delta V=\Delta V_p\equiv2pU$.  At resonance, $N-p$ particles tunnel between wells.
  The insert is a zoom around $\Delta V/2U = 2$.\label{fig:3}}
\end{figure}

All particles occupy the right well with probability
\begin{equation}
  P_0(t) = 1 - P_N(t) = \cos^2(\omega_Nt/2),
\end{equation}
In Fig. \ref{fig:3}(a), we
plot the probability densities $P_{n_L}(t)$ and the average occupation
$\n(t)$ as a function of time. The tunneling period is
$T_N\equiv2\pi/\omega_N$. The average occupation and variance are
\begin{align}
  \n(t) &= N\sin^2(\omega_Nt/2),\\
  \vn(t) &= (N^2/4)\sin^2(\omega_Nt).
\end{align}
In this regime, as in the noninteracting case, all $N$ particles oscillate
sinusoidally between wells.  There are two important differences.  The
first is that the period of oscillation depends on $N$ and can become quite large for large values of $N$. Note that, for large $N$, self-trapping is observed for exponentially long times, in agreement with mean-field approaches.  Second, at time
$t = T_N/4$, we find that $P_N=P_0=1/2$.  At this time, all particles
simultaneously occupy both wells and the system is described by a NOON-like MS state, a new prediction that cannot be achieved in the mean-field limit, and which is intriguing in the framework of quantum information.

Guided by this interest, let us characterize the  entanglement at $t=T_N/4$, by utilizing four
standard entanglement measures: the average local impurity, or ``Q-measure''~\cite{Brennen:2003}, the local entropy of
entanglement~\cite{Bennett:1996}, the Schmidt rank
$k$~\cite{Terhal:2000}, and a ``macroscopic superposition size'' (MSS) based on physical measurement~\cite{Korsbakken:2007}.  The $Q$-measure is given by $Q
=[(N+1)/N][1-(\Tr\rho_L^2+\Tr\rho_R^2)/2]$, where
$\rho_{L(R)}=\Tr_{R(L)}|\psi(t)\rangle\langle\psi(t)|$.  The entropy
is $S = -\sum_{n_L=0}^NP_{n_L}(t)\log_{N+1}P_{n_L}(t)$, and the
Schmidt rank $k$ is given by the number of non-zero eigenvalues of
the reduced density matrix $\rho_L$. Finally, the MSS measure is $C_{\delta}=N/n_{\mathrm{min}}$, where $n_{\mathrm{min}}$ is the minimum number of particles one must to measure to distinguish both branches, i.e., $|N-n_L,\,n_L\rangle$ and $|n_L,\,N-n_L\rangle$.  These measures take the value $Q=S=0$, $k=1$, and $C_{\delta}=0$ if and
only if $|\psi(t)\rangle$ is a pure state.  This occurs at $t=
T_N/2$ and $T_N$.  At time $T_N/4$, we find that each measure
reaches a maximum value of $Q=N/[2(N+1)]$, $S=\log_{N+1}(2)$,
$k=2$, and $C_{\delta}=N/(n_L+1)$.

We proceed to consider the effects of tilt or bias in the LMG model.  For $\zeta \ll 1$ the tunneling between wells is extremely sensitive to tilt $\Delta V$.   Using perturbation theory, it was shown that when $\Delta V > 2\Delta E_N/N$ the eigenstates are nearly perfect number-states of the form $|n_L,\,N-n_L\rangle$ and $|N-n_L,\,n_L\rangle$.   In this case, since the number states are near-eigenstates of the Hamiltonian, the initial condition $|\psi(0)\rangle$ is nearly stationary and
tunneling between wells is strongly suppressed.  This is quite
different from suppression of tunneling due to thermal effects or
coupling to a reservoir~\cite{Paz:1993,pitaevskii2001,Antunes:2006};
our system is closed and suppression is due to an internal
parameter, namely, imperfections in the trapping  potential.

\begin{figure}\center
\includegraphics[width=3.25in]{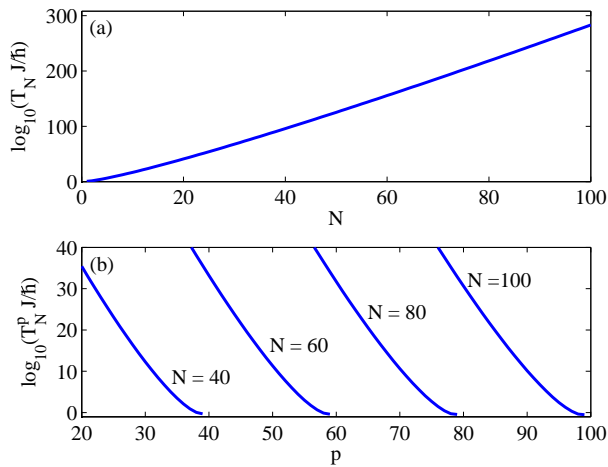}
\caption{(Color online) {\it Tunneling periods in a many-body system.} (a) Tunneling period $T_N$ versus the total number of particles $N$
  when $\zeta = 0.1$ and $\Delta V=0$.  For large $N$, tunneling becomes
  very slow.  (b) At resonance, $\Delta V = \Delta V_p \equiv 2pU$, only $N-p$
  particles tunnel between wells. Shown are the tunneling periods
  $T_N^p$ versus $p$ for $N = 40$ to 100 with $\zeta =
  0.1$.  At resonance, the oscillations can be hundreds of orders of
  magnitude faster than in the symmetric case.}
\label{fig:5}
\end{figure}

However, for stronger tilts quasi-degenerate MS eigenstates of the form $|\phi_{\pm};p\rangle \equiv\left(|N-p,\,p\rangle \pm |0,\,N\rangle\right)/ \sqrt{2}$  reappear~\cite{Dounas-Frazer:2007} for $\Delta V = \Delta V_p $, with a splitting in the energies between the states $|\phi_{\pm};p\rangle$ equal to $ \Delta E_N^p$, where
\begin{equation}
  \Delta E_N^p = \frac{4U(\zeta/2)^{N - p}(N-p)}{(N - p - 1)!}
    \sqrt{\bino{N}{p}},
    \label{eqn:splitting}
\end{equation}
In this case, the potential difference is exactly compensated by the repulsive interaction of $p$ particles in the lower well. Then, at resonance, the tunneling frequency is $\omega_N^p = \Delta
E_N^p/\hbar.$
The average
occupation of the left well is
\begin{equation}
  \n(t) = (N-p)\sin^2(\omega_N^pt/2)
\end{equation}
to lowest order in $\zeta$.  Here, $N-p$ particles tunnel between wells with period
$T_N^p=2\pi/\omega_N^p$.  At time $t=T_N^p/2$, $N-p$ particles are
in the left well. To compensate the tilt, $p$ particles remain in
the right well at all times.  When $t=T_N^p/4$, the system is
described by an MS state such that $P_p=P_0=1/2$ and the
entanglement measures $Q$, $S$, and $k$ reach the same values as in
the symmetric case.

The entanglement measured by $C_{\delta}$ is again $C_{\delta}=N$, provided that the measure is taken in the higher well. Interestingly enough, if the measure is taken in the lower one, the entanglement measured is smaller,  $C_{\delta}=N/(p+1)$, since $p+1$ particles must be measured to distinguish between both branches.  In Fig. \ref{fig:3}(b) the tunneling
dynamics for the second resonance, i.e., $p=2$,  are illustrated for a
system of $N=7$ particles. Near a resonance, tunneling is suppressed
when
\begin{equation}
  |\Delta V-\Delta V_p| > 2\Delta E^p_N/(N-p).
\end{equation}
This is due to the fact that, near the resonance, the eigenstates are again near-perfect number states when the deviations of the tilt with respect to the resonant one exceed this quantity~\cite{Dounas-Frazer:2007}.  Then, the condition for the suppression of the  tunneling in the symmetric case is that the tilt exceeds $ 2\Delta E_N/N$ while in the case of the asymmetric potential, when the tilt coincides with a resonance, the condition for the suppression of the tunneling is that the difference between the tilt and that of the resonance exceeds $ 2\Delta E^p_N/(N-p)$.

Moreover, tunneling near resonance is much faster than tunneling in a symmetric potential since both tunneling frequencies, $\omega_N^p$ and  $\omega_N $, are proportional to  $  \Delta E_N^p/\hbar$ and  $ \Delta E_N/\hbar$ respectively, and we have shown that  $\Delta E_N^p$ is greater than $\Delta E_N$ by many orders of magnitude.  In Fig.~\ref{fig:5}(a), we show the symmetric tunneling period $T_N$ versus
$N$ when $\zeta = 0.1$. Clearly, $T_N$ becomes very long as $N$
becomes large. For instance, in a typical symmetric double-well used
in experiments \cite{anderlini2007}, 200 \Rb atoms tunnel
between wells with period $T_{200}=1.15\times10^{635}$~ms when $\zeta
= 0.0964$.  Furthermore, tunneling is completely suppressed for
deviations in the tilt greater than $4.16\times10^{-636}\nK$.
Obviously, one does not expect to observe many-body tunneling in
this regime. Notice that, under the same conditions, systems with as few as
$N=1$, 2, and 3 \Rb atoms yield tunneling times as long as
$T_1=466$~ms, $T_2 = 4840$~ms, and $T_3 = 134000$~ms, respectively.
Even in a few-particle system, tunneling times can be prohibitively
long, demonstrating self-trapping behavior~\cite{Albiez:2005}.

\begin{figure}\center
\includegraphics[width=0.48\textwidth]{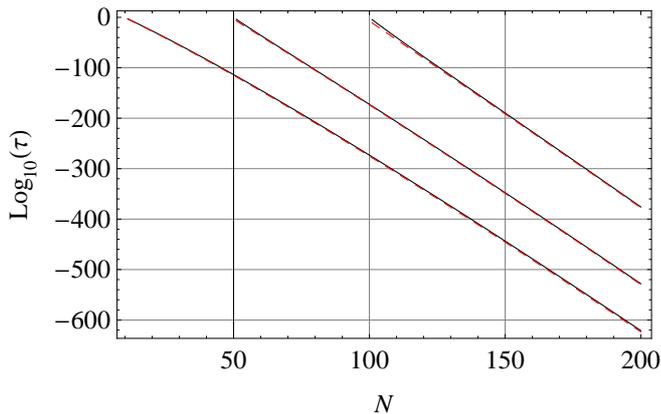}
\caption{\textit{Embedding the MS state in the many body wavefunction.}  Relative increase $\tau$ in tunneling time and robustness against potential fluctuations, for an MS state of $N-p=10$, 50, 100 particles (left to right); plotted as a function of the number of particles in the full state, $N$, for exact (solid black curves) and approximate (dashed red curves) expressions, all for $\zeta=0.1$.
\label{fig:4}}
\end{figure}

However, tunneling at resonance can be hundreds of orders of
magnitude faster than the symmetric case, as in Fig. \ref{fig:5}(b).
For the 200-atom system discussed above, when $p=197$, we find  that
$N-p=3$ particles tunnel between wells with period
$T_{200}^{197}=117$ ms.  This resonance occurs when $\Delta V=\Delta
V_{197} = 210\nK$.  An MS state of the form
$|\psi\rangle=(|3,\,197\rangle-i|0,\,200\rangle)/ \sqrt{2}$ will be
realized at $T_{200}^{197}/4 = 29.25$ ms.  Likewise, we find that
$T_{200}^{198}=34.3$  ms and $T_{200}^{199} = 33.0$ ms when $\Delta
V = \Delta V_{198} = 211\nK$ and $\Delta V_{199} = 212\nK$,
respectively. Finally, at resonance, this system is sensitive to deviations
in the tilt on the order of $0.273\nK$, $1.40\nK$, and $2.90\nK$ for
$p=197$, 198, and 199, respectively. Thus, the observation of the
tunneling of a few \Rb  atoms is made possible by tunneling
resonances in a many-body system.  Moreover, embedding NOON-like MS states in the many body wavefunction leads to the possibility of larger MS states of more than three particles; the scaling of the advantage in tunneling time gained via tilt can be calculated as $\tau\equiv \Delta E_N/\Delta E_N^p$, from Eqs.~(\ref{eqn:splittingNoTilt}) and~(\ref{eqn:splitting}).  Taking $p'\equiv N-p$ as the number of atoms in the embedded MS state, and using Stirling's approximation $\ln n!\simeq n\ln n - n$, we find \begin{eqnarray}\ln \tau&\simeq& \textstyle(-\ln n + 1+ \ln\frac{\zeta}{2})n +(\frac{p'}{4n}+\frac{p'^2}{12 n^2}-\frac{1}{2}\ln n)p'\nonumber\\
&&\textstyle  + (\frac{3}{2}\ln p' -\frac{3}{2}+\frac{1}{2}\ln 4-\ln \zeta)p'\,;\end{eqnarray} the second set of cross terms is vital.  This expansion matches the exact expression very well, as illustrated in Fig~\ref{fig:4}.

In conclusion, we used the two-mode approximation to develop a Fock
space picture of a system of ultracold bosons in a tilted two-well
potential, covering all regimes of barrier height, from the Josephson regime to the regime in which one can find NOON states.  In the latter regime, which occurs when the barrier is high, a small tilt causes the complete suppression of
tunneling, leading to self-trapping.  Long tunneling times prevent the
observation of many-body tunneling even in a symmetric potential and MS states are too sensitive to fluctuations in the trapping potential to be realistic.
However, in this regime, tunneling resonances occur when the
tilt can be compensated by atom-atom interactions.  At resonance,
tunneling is much faster and less sensitive to tilt than in a
symmetric potential.  Furthermore, tunneling resonances can be used to
create NOON-like MS states embedded in and protected by a larger many-body system.

We thank Ignacio Cirac, Ann Hermundstad, William Phillips, and Trey
Porto for useful discussions.  This work was supported by the National Science Foundation under Grant PHY-0547845 as part of the NSF CAREER program,  by the Fulbright Commission, by Spain's Ministerio de Educaci\'on y Ciencia (MEC), and by the Fundaci\'on Espa\~nola de Ciencia y Tecnolog\'{\i}a (FECYT).

%\bibliographystyle{eplbib}
%\bibliography{referencias}

\begin{thebibliography}{10}
\expandafter\ifx\csname url\endcsname\relax\def\url#1{\texttt{#1}}\fi

\bibitem{lewensteinM2007}
\Name{Lewenstein, M.~Sanpera A., Ahufinger V., Damski B., Sen~De A. \and Sen
  U.} \REVIEW{Adv. Phys. }{56}{2007}{243}.

\bibitem{bloch2008}
\Name{Bloch I., Dalibard J. \and Zwerger W.} \REVIEW{Rev. Mod. Phys.
  }{80}{2008}{885}.

\bibitem{Shin:2005}
\Name{Shin Y., Jo G.-B., Saba M., Pasquini T.~A., Ketterle W. \and Pritchard
  D.~E.} \REVIEW{Phys. Rev. Lett.}{95}{2005}{170402}.

\bibitem{Albiez:2005}
\Name{Albiez M., Gati R., F{\"o}lling J., Hunsmann S., Cristiani M. \and
  Oberthaler M.} \REVIEW{Phys. Rev. Lett.}{95}{2005}{010402}.

\bibitem{Tolra:2004}
\Name{Tolra B.~L., O'Hara K.~M., Huckans J.~H., Phillips W.~D., Rolston S.~L.
  \and Porto J.~V.} \REVIEW{Phys. Rev. Lett.}{92}{2004}{190401}.

\bibitem{lipkinHJ1965}
\Name{Lipkin H.~J., Meshkov N.~\and Glick A.~J.} \REVIEW{Nucl. Phys.
  }{62}{1965}{188}.

\bibitem{vidalJ2004}
\Name{Vidal J., Palacios G. \and Aslangul C.} \REVIEW{Phys. Rev. A
  }{70}{2004}{062304}.

\bibitem{Dagnino:2009}
\Name{Dagnino D., Barber\'an N., Lewenstein M. \and Dalibard J.} \REVIEW{Nature Phys. }{5}{2009}{431}.

\bibitem{leggett2001}
\Name{Leggett A.~J.} \REVIEW{Rev. Mod. Phys. }{73}{2001}{307}.

\bibitem{Milburn:1997}
\Name{Milburn G.~J., Corney J., Wright E.~M. \and Walls D.~F.} \REVIEW{Phys. Rev. A}{55}{1997}{4318}.

\bibitem{Smerzi:1997}
\Name{Smerzi A., Fantoni S., Giovanazzi S. \and Shenoy S.~R.} \REVIEW{Phys. Rev. Lett.}{79}{1997}{4950}.

\bibitem{Zapata:1998}
\Name{Zapata I., Sols F. \and Legget A.~J.} \REVIEW{Phys. Rev. A
  }{57}{1998}{R28}.

\bibitem{masiello2006}
\Name{Masiello D., McKagan S.~B. \and Reinhardt W.~P.} \REVIEW{Phys. Rev. A
  }{72}{2006}{063624}.

\bibitem{Alon:2008}
\Name{Alon O.~E., Streltsov A.~I. \and Cederbaum L.~S.} \REVIEW{Phys. Rev. A
  }{77}{2008}{033613}.

\bibitem{Dounas-Frazer:2007}
\Name{Dounas-Frazer D.~R., Hermundstad A.~M. \and Carr L.~D.} \REVIEW{Phys. Rev. Lett.}{99}{2007}{200402}.

\bibitem{watanabeG2010}
\Name{G. Watanabe} \REVIEW{Phys. Rev. A}{81}{2010}{021604(R)}.

\bibitem{weissC2008}
\Name{Weiss C. \and Teichmann N.} \REVIEW{Phys. Rev. Lett.}{100}{2008}{140408}.

\bibitem{friedman2000}
\Name{Friedman J.~R., Patel V., Chen W., Tolpygo S.~K. \and Lukens J.~E.}
  \REVIEW{Nature }{406}{2000}{43}.

\bibitem{Dounas-Frazerthesis:2006}
\Name{Dounas-Frazer D.~R.} \Book{Ultracold bosons in a multi-dimensional,
  tilted, double-well trap: Potential decoherence, tunneling resonances, and
  two-level phenomena} Master's thesis, Colorado School of Mines (2007).

\bibitem{Salgueiro:2006}
\Name{Salgueiro A.~N., de~Toledo~Piza A.~F.~R., Lemos G.~B., Drumond R., Nemes
  M.~C. \and Weidem{\"u}ller M.} \REVIEW{Eur. Phys. J. D}{44}{2007}{537}.

\bibitem{Paz:1993}
\Name{Paz J.~P., Habib S. \and Zurek W.~H.} \REVIEW{Phys. Rev. D
  }{47}{1993}{488}.

\bibitem{pitaevskii2001}
\Name{Pitaevskii L. \and Stringari S.} \REVIEW{Phys. Rev. Lett.
  }{87}{2001}{180402}.

\bibitem{Antunes:2006}
\Name{Antunes N.~D., Lombardo F.~C., Monteoliva D. \and Villar P.~I.}
  \REVIEW{Phys. Rev. E }{73}{2007}{066105}.

\bibitem{Mahmud:2005}
\Name{Mahmud K., Perry H. \and Reinhardt W.} \REVIEW{Phys. Rev. A
  }{71}{2005}{023615}.

\bibitem{Huang:2006}
\Name{Huang Y.~P. \and Moore M.~G.} \REVIEW{Phys. Rev. A }{73}{2006}{023606}.

\bibitem{anderlini2007}
\Name{Anderlini M., Lee P.~J., Brown B.~L., Sebby-Strabley J., Phillips W.~D.
  \and Porto J.~V.} \REVIEW{Nature }{448}{2007}{452}.

\bibitem{foelling2007}
\Name{Foelling S., Trotzky S., Cheinet P., Feld M., Saers R., Widera1 A.,
  Muller T. \and Bloch1 I.} \REVIEW{Nature }{448}{2007}{1029}.

\bibitem{Javanainen:1987}
\Name{Javanainen J.} \REVIEW{Phys. Rev. Lett.}{57}{1987}{3164}.

\bibitem{Ananikian:2006}
\Name{Ananikian D. \and Bergeman T.} \REVIEW{Phys. Rev. A
  }{73}{2006}{013604}.

\bibitem{Kohler:2001}
\Name{Kohler S. \and Sols F.} \REVIEW{Phys. Rev. A }{63}{2001}{053605}.

\bibitem{Paraoanu:2001}
\Name{Paraoanu G.~S., Kohler S., Sols F. \and Legget A.~J.} \REVIEW{J. Phys. B: At. Mol.
  Opt. Phys. }{34}{2001}{4689}.

\bibitem{Leggett:1991}
\Name{Legget A.~J. \and Sols F.} \REVIEW{Found. Phys. }{21}{1991}{353}.

\bibitem{Sakmann:2009}
\Name{Sakmann K., Streltsov A.~I., Alon O.~E. \and Cederbaum L.~S.}
  \REVIEW{Phys. Rev. Lett.}{103}{2009}{220601}.

\bibitem{Gati:2006}
\Name{Gati R., Esteve J., Hemmerling B., Ottenstein T.~B., Appmeier J., Weller
  A. \and Oberthaller M.~K.} \REVIEW{New J. Phys.}{8}{2006}{189}.

\bibitem{Jaaskelainen:2005}
\Name{{J\"a\"askel\"ainen} M. \and Meystre P.} \REVIEW{Phys. Rev. A
  }{71}{2005}{043603}.

\bibitem{Ferrari:2006}
\Name{Ferrari G., Poli N., Sorrentino F. \and Tino G.~M.} \REVIEW{Phys. Rev. Lett.}{97}{2006}{060402}.

\bibitem{Hall:2006}
\Name{Hall B.~V., Whitlock S., Anderson R., Hannaford P. \and Sidorov A.~I.}
  \REVIEW{Phys. Rev. Lett.}{98}{2007}{030402}.

\bibitem{Seaman:2006}
\Name{Seaman B.~T., Kr{\"a}mer M., Anderson D.~Z. \and Holland M.~J.}
  \REVIEW{Phys. Rev. A }{75}{2007}{023615}.

\bibitem{Brennen:1999}
\Name{Brennen G.~K., Caves C.~M., Jessen P.~S. \and Deutsch I.~H.}
  \REVIEW{Phys. Rev. Lett.}{82}{1999}{1060}.

\bibitem{Zapata:2003}
\Name{Zapata I., Sols F. \and Legget A.~J.} \REVIEW{Phys. Rev. A
  }{67}{2003}{021603}.

\bibitem{Brennen:2003}
\Name{Brennen G.~K.} \REVIEW{Quant. Inf. Comp.}{3}{2003}{619}.

\bibitem{Bennett:1996}
\Name{Bennett C.~H., Bernstein H.~J., Popescu S. \and Schumacher B.}
  \REVIEW{Phys. Rev. A }{53}{1996}{2046}.

\bibitem{Terhal:2000}
\Name{Terhal B.~M. \and Horodecki P.} \REVIEW{Phys. Rev. A
  }{61}{2000}{040301}.

\bibitem{Korsbakken:2007}
\Name{Korsbakken J.~I., Whaley K.~B., Dubois J. \and Cirac J.~I.}
  \REVIEW{Phys. Rev. A }{75}{2007}{042106}.

\end{thebibliography}

\end{document}